\documentclass[final,aps,prb,twocolumn,superscriptaddress,showpacs]{revtex4-1}

\usepackage{graphicx}
\usepackage{dcolumn}
\usepackage{amsmath,bm}
\usepackage{color}%
\usepackage{multirow}
\usepackage{array} 
\usepackage{xr}

\newcommand*{\rom}[1]{\expandafter\@slowromancap\romannumeral #1@}
\makeatletter
\newcommand*{\addFileDependency}[1]{
  \typeout{(#1)}
  \@addtofilelist{#1}
  \IfFileExists{#1}{}{\typeout{No file #1.}}
}
\makeatother
 
\newcommand*{\myexternaldocument}[1]{%
    \externaldocument{#1}%
    \addFileDependency{#1.tex}%
    \addFileDependency{#1.aux}%
}

\myexternaldocument{SI}

\begin{document}

\title{Overcoming Quantum Resistivity Scaling in Nanoscale Interconnects Using Delafossite PdCoO$_2$}

\author{Seoung-Hun Kang}
\thanks{These two authors contributed equally.}
\thanks{Corresponding author: physicsksh@gmail.com}
\affiliation{Department of Information Display, Kyung Hee University, Seoul 02447, Korea}
\affiliation{Department of Physics, and Research Institute for Basic Sciences, Kyung Hee University, Seoul 02447, Korea}
\affiliation{Materials Science and Technology Division, Oak Ridge National Laboratory, Oak Ridge, TN 37831, USA}

\author{Youngjun Lee}
\thanks{These two authors contributed equally}
\affiliation{Department of Physics, and Research Institute for Basic Sciences, Kyung Hee University, Seoul 02447, Korea}

\author{Sangmoon Yoon}
\affiliation{Department of Physics, Gachon University, Seongnam 13120, Korea}

\author{Jong Mok Ok}
\affiliation{Department of Physics, Pusan National University, Busan 46241, Korea}

\author{Mina Yoon}
\affiliation{Materials Science and Technology Division, Oak Ridge National Laboratory, Oak Ridge, TN 37831, USA}

\author{Ho Nyung Lee}
\affiliation{Materials Science and Technology Division, Oak Ridge National Laboratory, Oak Ridge, TN 37831, USA}

\author{Young-Kyun Kwon}
\thanks{Corresponding author: ykkwon@khu.ac.kr}
\affiliation{Department of Information Display, Kyung Hee University, Seoul 02447, Korea}
\affiliation{Department of Physics, and Research Institute for Basic Sciences, Kyung Hee University, Seoul 02447, Korea}

\date{\today}

\begin{abstract}
Continued scaling into the sub-7 nm regime exacerbates quantum-limited resistivity in Cu interconnects. We evaluated layered PdCoO$_2$ and explicitly benchmarked it against Cu to identify mechanisms that maintain conductivity under confinement. Using a momentum-resolved relaxation-time formalism derived from the conductivity tensor, we link k- and energy-resolved velocities, lifetimes, and mean-free-paths (MFPs) to thickness-dependent resistivity for films and wires. PdCoO$_2$ exhibits quasi-2D transport with high in-plane velocities and strongly anisotropic MFPs ($\sim 15\,\mathrm{nm}$ in-plane, $\sim 3\,\mathrm{nm}$ out-of-plane near E$_F$), whereas Cu shows an isotropic $\sim 22\,\mathrm{nm}$ MFP. Under identical boundary conditions—including a realistic 2 nm liner/diffusion barrier for Cu—PdCoO$_2$ displays suppressed boundary scattering and a much slower resistivity increase from bulk down to sub-30 nm, preserving near-bulk conductivity and remaining viable at ~2 nm. Thickness trends reveal dual slope changes in PdCoO$_2$ ($\sim 35\,\mathrm{nm}$ and $\sim 7\,\mathrm{nm}$) set by anisotropic MFPs, contrasting with Cu’s single characteristic scale ($\sim 40\,\mathrm{nm}$). The calculated bulk values and scaling curves track available measurements for both materials. These results establish PdCoO$_2$ as a scalable interconnect that outperforms Cu under quantum confinement and provide a quantitative framework to screen layered conductors for next-generation nanoelectronic interconnects.

\end{abstract}

\keywords{quantum effects, interconnect resistivity, PdCoO$_2$, delafossite materials, ultra-thin electronic devices, first-principles calculation} 

\maketitle

\section{\label{sec:level1}Introduction}
The relentless miniaturization of semiconductor devices toward dimensions below 5 nm has significantly improved computing performance and integration density, driving remarkable advances in electronic technologies~\cite{Radamson2024}. However, as device dimensions approach atomic scales, quantum-mechanical phenomena increasingly dominate electron transport, introducing critical new challenges~\cite{Radamson2024,Badawy2024}. In particular, traditional interconnect materials such as copper, which exhibit outstanding electrical conductivity at bulk scales, experience a pronounced increase in resistivity at these ultra-thin dimensions due to intensified electron-boundary scattering, grain-boundary effects, and drastically reduced electron mean free paths~\cite{Pan2014, Graham2010, Gall2020}. This increase in resistivity severely undermines the signal integrity, power efficiency, and overall performance of nanoscale circuits, presenting a significant barrier to further technological advancement~\cite{Moon2023}.

Moreover, copper interconnect technology necessitates additional structural complexity, including diffusion barriers and adhesion liners to prevent metal migration into adjacent dielectric layers. The insertion of these barrier and liner materials significantly reduces the effective conductive cross-sectional area of the interconnect, exacerbating resistivity problems and complicating the integration process~\cite{Steinhogl2004}. Consequently, the search for alternative materials with inherently superior electrical properties at the nanoscale, as well as reduced integration complexity, has become essential to sustain future semiconductor scaling trends.

In this context, delafossite oxide materials, particularly PdCoO$_2$, have recently emerged as exceptionally promising candidates to replace traditional metallic interconnects~\cite{JS2024,Harada2025}. Delafossite oxides possess a unique layered crystal structure characterized by alternate metallic and oxide planes, facilitating remarkable electronic transport properties~\cite{Shannon1971, Mackenzie2017}. PdCoO$_2$, in particular, has garnered considerable attention due to its extraordinary electrical conductivity, rivaling or even exceeding that of conventional metals such as copper and silver. Furthermore, PdCoO$_2$ exhibits a strongly anisotropic, quasi-two-dimensional (2D) Fermi surface structure, which promotes efficient, directionally confined electron transport parallel to its atomic layers~\cite{Eyert2008, Mackenzie2017}.

Unlike graphene, another prominent 2D conductor, which suffers from relatively low carrier densities that limit its practical applicability in interconnect technology~\cite{Wu2017}, PdCoO$_2$ uniquely combines a high carrier density with exceptional electron mobility. This advantageous combination not only sustains a superior electrical conduction at reduced dimensions but also significantly mitigates the detrimental impact of quantum confinement effects. Furthermore, PdCoO$_2$ intrinsically incorporates stable oxide layers, eliminating the need for separate diffusion barriers or liners. As a result, its adoption can maximize the effective conductive area, reduce interconnect complexity, and maintain a lower overall resistivity within integrated circuits, greatly simplifying device fabrication processes. 

Recent advances in thin-film deposition methods, such as molecular beam epitaxy (MBE), pulsed laser deposition (PLD), and chemical vapor deposition (CVD) have successfully realized high-quality PdCoO$_2$ thin films, maintaining their bulk-like conductivity and mechanical stability down to ultrathin dimensions~\cite{Brahlek2019, Harada2018}. A similar class of delafossites, such as PdCrO$_2$ and CuCrO$_2$ have been also sucessfully demonstrated by PLD~\cite{OkCuCrO2,OkPd}. Despite these achievements, critical open questions remain about precisely how thickness reduction affects the quantum-scale resistivity of PdCoO$_2$, particularly regarding the interplay between electron-phonon scattering and reduced electron momentum relaxation pathways at ultrathin dimensions.

In this study, we systematically investigate the thickness-dependent electrical resistivity of PdCoO$_2$ using comprehensive first-principles computational methods. Specifically, we rigorously quantify electron-phonon coupling and momentum relaxation times across different electronic bands and momentum states at nanoscale thicknesses, enabling detailed insight into quantum mechanical scattering mechanisms. Our results reveal that PdCoO$_2$ retains outstanding electrical conductivity even at thicknesses below 5 nm, significantly exceeding copper and other conventional metallic interconnect materials. This robustness arises from the layered quasi-2D transport in PdCoO$_2$---high in-plane Fermi velocities and anisotropic mean free paths that suppress boundary scattering along the conducting planes---together with the absence of liner/barrier penalties required for Cu, yielding slower resistivity scaling and bulk-like conductivity down to the few-nanometer regime. This demonstrates the potential of PdCoO$_2$ to overcome the existing nanoscale resistivity bottleneck, thus positioning it as an enabling breakthrough material for continued CMOS scaling and the development of next-generation nanoscale electronic devices.

\section{\label{sec:level1}Computational details}

First-principles calculations were performed based on density functional theory (DFT)\cite{Kohn1965, Hohenberg1964}, as implemented in Quantum ESPRESSO\cite{Giannozzi2009}. The plane-wave basis set utilized a kinetic-energy cutoff of 80 Ry, with norm-conserving pseudopotentials and the Perdew-Burke-Ernzerhof (PBE) generalized gradient approximation~\cite{Perdew1996} for the exchange-correlation functional. Self-consistent ground-state calculations were conducted on a rhombohedral lattice (lattice constant: 6.1535 \AA), employing a uniform $12\times12\times12$ $k$-point mesh corresponding to a reciprocal-space sampling density of 0.214 \AA$^{-1}$.

Phonon calculations were performed using density functional perturbation theory (DFPT)~\cite{Baroni2001} with a $4\times4\times4$ $q$-point grid. Electron-phonon (e-ph) interaction matrix elements were subsequently interpolated using maximally localized Wannier functions implemented in the EPW package~\cite{Pizzi2020,Ponce2016}, employing finely converged grids of $60\times60\times60$ for $k$-points and 10,000 points for $q$-space sampling (see SI-Fig.S1)~\cite{Jamal2016}. Electrical resistivity was evaluated by solving the Boltzmann transport equation using the BoltzTraP code~\cite{Madsen2006}. Finally, to quantify thickness-dependent resistivity effects in thin films and wires, we utilized our previously developed momentum-resolved relaxation-time approach~\cite{Lee2025}.
 
\section{\label{sec:level1}Results and discussion}

\begin{figure}[t]
\includegraphics[width=1\linewidth]{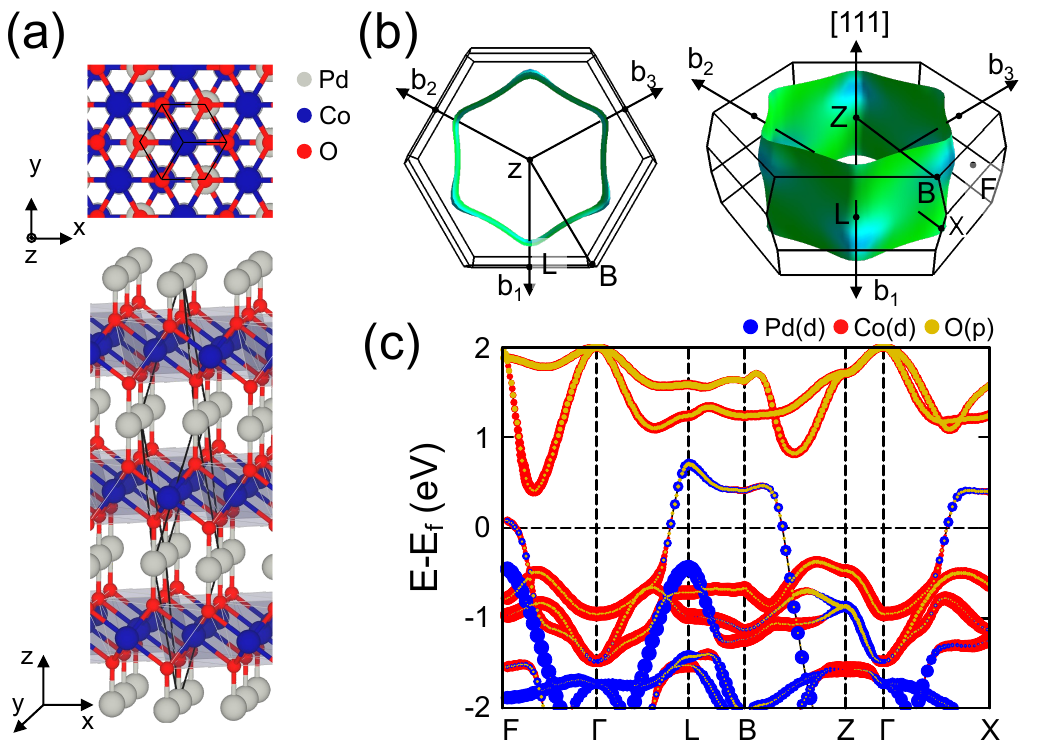}
\caption{Crystal structure and electronic properties of PdCoO$_2$. (a) Top and side views of the bulk PdCoO$_2$ crystal structure. PdCoO$_2$ adopts a rhombohedral delafossite structure composed of alternating metallic Pd layers and insulating CoO$_2$ layers stacked along the [111] crystallographic direction. The unit cell, indicated by the black box, highlights the hexagonal arrangement of Pd atoms that form conductive two-dimensional (2D) planes, confining electron transport within these layers.
(b) Calculated Fermi surface of bulk PdCoO$_2$, clearly illustrating its pronounced 2D cylindrical geometry. The Fermi surface shows negligible dispersion along the [111] direction, confirming highly anisotropic electronic transport and indicating effective confinement of electrons to the Pd planes. $z$-axis is the direction of [111] in the rhombohedral delafossite structure. This 2D confinement significantly enhances electron mobility within the layers.
(c) Orbital-projected electronic band structure of PdCoO$_2$, detailing the contributions from Pd $d$ (blue), Co $d$ (red), and O $p$ (orange) orbitals. The dominant contribution of Pd $d$ orbitals near the Fermi level demonstrates their central role in the superior conductivity, while the Co $d$ and O $p$ states primarily lie away from the Fermi energy, reinforcing the insulating character of the CoO$_2$ layers and further ensuring high anisotropic conductivity within the metallic Pd layers.
\label{Fig1}}
\end{figure}
Using the optimized PdCoO$_2$ crystal structure presented in Fig.~\ref{Fig1}(a), we analyzed the electronic structure and transport characteristics crucial for its potential application in nanoscale electronics. PdCoO$_2$ crystallizes in a rhombohedral delafossite structure, consisting of alternating metallic Pd layers and insulating CoO$_2$ layers stacked along the [111] crystallographic direction. Within this structure, Pd and Co atoms form hexagonal close-packed (HCP) lattices exhibiting a six-fold symmetry derived from triangular sublattices stacked with slight relative shifts. Each metal atom coordinates with six nearest neighbors, resulting in distorted octahedral geometries, whereas oxygen atoms occupy octahedral interstitial sites, forming distorted hexagonal prismatic environments around the metal centers. Weak van der Waals interactions predominantly govern the interlayer bonding, generating significant structural anisotropy and confining electron transport predominantly within the metallic Pd atomic planes.

The two-dimensional (2D) nature of electronic transport in PdCoO$_2$ is further emphasized by the calculated Fermi surface shown in Fig.~\ref{Fig1}(b). The Fermi surface exhibits a cylindrical geometry aligned along the [111] stacking direction, underscoring negligible electronic dispersion out-of-plane. This feature indicates weak interlayer electronic coupling and confirms intrinsic anisotropic conductivity, which is beneficial for high electron mobility and enhanced in-plane conduction.

The calculated electronic band structure, presented in Fig.~\ref{Fig1}(c), reveals additional insights into the material's transport behavior. Electronic states near the Fermi level predominantly originate from partially filled Pd and Co $d$ orbitals, with only minor contributions from O $p$ orbitals. The metallic character of PdCoO$_2$ arises primarily due to these $d$ bands, which exhibit strong orbital hybridization. A small energy gap separates these bands from the underlying valence states of O $p$. Importantly, the steep dispersion observed near the Fermi level indicates high carrier velocities and suggests excellent intrinsic electron mobility, underscoring the PdCoO$_2$'s potential as a highly efficient charge conductor for next-generation nanoscale electronic and interconnect applications.

\begin{figure}[t]
\includegraphics[width=1\linewidth]{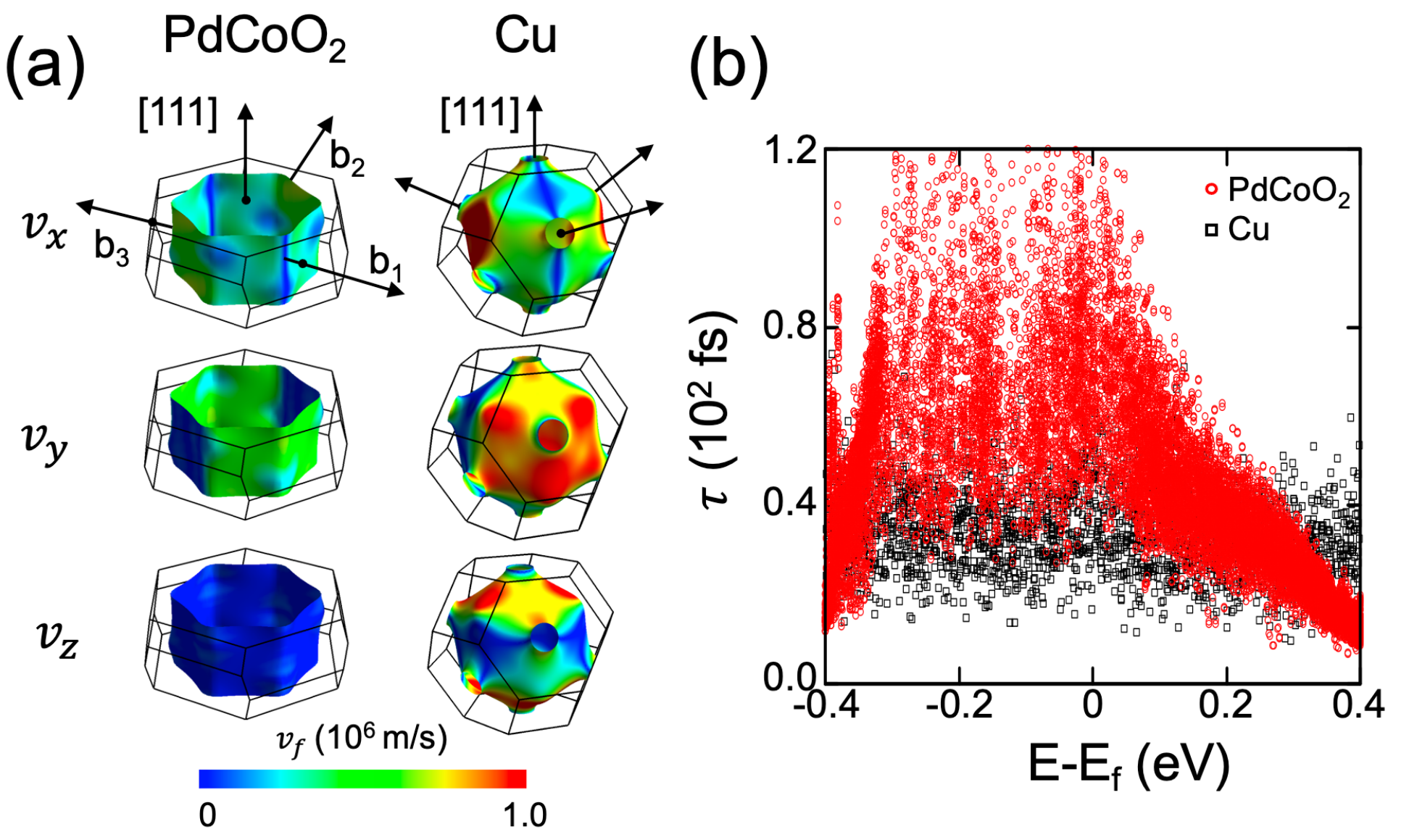}
\caption{Fermi velocity and electron-phonon relaxation times in PdCoO$_2$ and Cu.
(a) Color-mapped Fermi surfaces of PdCoO$_2$ and Cu, showing the directional distribution of Fermi velocities along the $x$, $y$, and $z$ axes. Red and blue colors indicate regions of high and low Fermi velocity, respectively. PdCoO$_2$ exhibits a highly anisotropic distribution with significantly higher Fermi velocities in the in-plane directions ($x$ and $y$) than those along the out-of-plane ($z$) direction, reflecting the quasi-two-dimensional nature of its electronic structure. In contrast, copper shows a nearly isotropic Fermi velocity, consistent with its three-dimensional, free-electron-like electronic behavior. (b) Energy-dependent electron relaxation times for PdCoO$_2$ and Cu at 300 K, estimated from the imaginary part of the electron self-energy using electron-phonon coupling. 
\label{Fig2}}
\end{figure}

Having established a comprehensive understanding of the fundamental electronic properties of PdCoO$_2$, we now investigate its thickness-dependent resistivity, a critical factor for nanoscale interconnect applications. To gain deeper insight into its intrinsic transport behavior, we focus on two key parameters: the Fermi velocity and the electron-phonon relaxation time. Both parameters directly influence the mean free path(MFP) of electrons, which governs the conductivity at reduced dimensions. For comparison, we include copper (Cu), the industry-standard interconnect material, which offers excellent bulk conductivity, but experiences significant performance degradation at nanometer scales~\cite{Pan2014, Graham2010, Gall2020}.
Figure~\ref{Fig2}(a) shows color-mapped Fermi surfaces of PdCoO$_2$ and Cu, with velocity components projected along the $x$, $y$, and $z$ directions. PdCoO$_2$ displays a strong anisotropy, with Fermi velocities in the plane ($x$, $y$) approximately 7.97 times higher than the out-of-plane ($z$) component. This anisotropic behavior results from its layered crystal structure, which confines conduction to the Pd atomic planes while suppressing interlayer electronic coupling.
In contrast, Cu exhibits a nearly isotropic Fermi surface with a uniform velocity distribution, consistent with its free-electron-like three-dimensional structure that supports omnidirectional charge transport. The directional velocity differences between the two materials are pronounced, reaching 0.48$\times$10$^6$~m/s along $x$, 0.56$\times$10$^6$~m/s along $y$, and 1.1$\times$10$^6$~m/s along $z$, further emphasizing their fundamentally different transport characteristics.
To assess the scattering behavior, we calculated the energy-dependent electron-phonon relaxation times $\tau(E)$ for both materials, as shown in Fig.~\ref{Fig2}(b). Both PdCoO$_2$ and Cu exhibit the peak relaxation time near the Fermi level, but PdCoO$_2$ shows a consistently higher values, averaging nearly twice those of Cu at 300 K. These extended lifetimes indicate reduced electron-phonon scattering and suggest longer coherence lengths for charge carriers in PdCoO$_2$.
In general, PdCoO$_2$ combines a high in-plane Fermi velocity with a long relaxation time, offering distinct advantages for directional charge transport. Although Cu remains effective in bulk due to its isotropic conductivity, PdCoO$_2$ is better suited for ultrathin interconnects, where planar conduction is dominant and quantum size effects intensify resistivity in conventional metals. These characteristics establish PdCoO$_2$ as a promising candidate for next-generation interconnects.

\begin{figure}[t]
\includegraphics[width=1\linewidth]{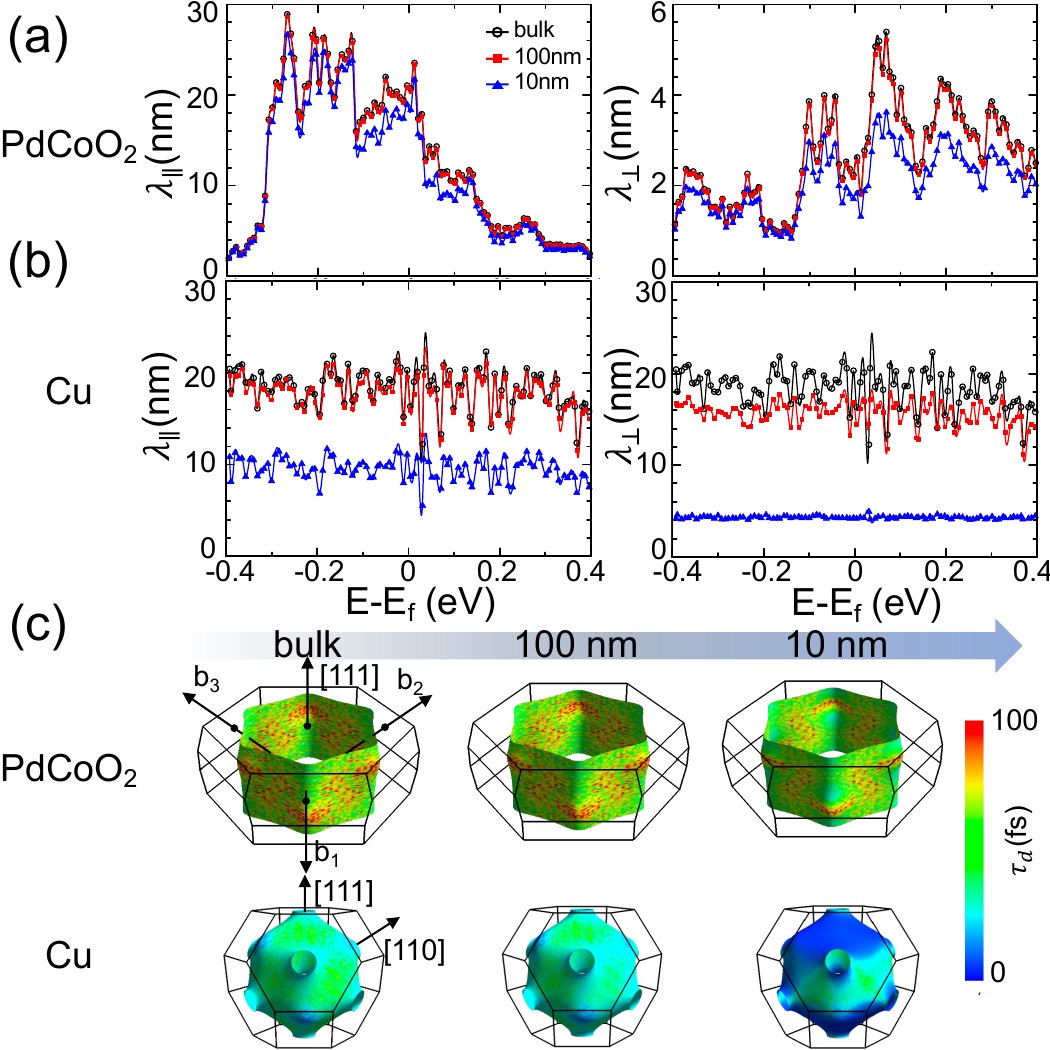}
\caption{Thickness-dependent electronic transport properties of PdCoO$_2$ and Cu.
(a) Thickness-dependent in-plane and out-of-plane mean free paths along the [111] direction near the Fermi level for thick bulk(black), 100~nm(red), and thin 10~nm films(blue) for PdCoO$_2$ and (b) Cu. The entire point represents the averaged mean free path at each energy described in SI-Fig.S3). Cu shows nearly isotropic and thickness-independent MFPs, unlike PdCoO$_2$, which exhibits a strong anisotropic reduction in MFP with decreasing thickness.
(c) Fermi surface projections of relaxation times for bulk, 100 nm, and 10 nm cases of PdCoO$_2$ and Cu. Blue and red colors indicate regions of low and high relaxation times, respectively. Thickness-dependent values for PdCoO$_2$ films were obtained using a momentum-resolved model based on bulk electron-phonon scattering data.
\label{Fig3}}
\end{figure}

We determined the MFP of bulk PdCoO$_2$ (see SI-Fig.S2) by combining the calculated Fermi velocity and electron–phonon relaxation time in Fig.~\ref{Fig2}. The MFPs of Cu and PdCoO$_2$ are comparable in magnitude near the Fermi level (see SI-Fig.S2).
In our momentum-resolved formalism, the MFP at each band index $n$ and $k$-point possesses a directional vector, which can be decomposed into two orthogonal components: the in-plane component ($\lambda\parallel$) perpendicular to the [111] direction and the out-of-plane component ($\lambda_\bot$) parallel to [111] (Fig.~\ref{Fig3}(a,b)). Based on this, we extracted the energy-resolved MFPs of PdCoO$_2$ and Cu for bulk and thin films with 100 and 10 nm in thickness and computed energy-averaged values for a direct comparison. 
As shown in Fig.~\ref{Fig3}(a), bulk PdCoO$_2$ exhibits a pronounced anisotropy near the Fermi level with an in-plane MFP of approximately 15 nm and an out-of-plane MFP of about 3 nm. This fivefold difference reflects the quasi-two-dimensional nature of PdCoO$_2$ consistent with its cylindrical Fermi surface. In contrast, bulk Cu exhibits an isotropic average over the directional components of the MFP,  with a value of approximately 22 nm regardless of the orientation, as shown in Fig.~\ref{Fig3} (b), approximately 1.5 times larger than the in-plane MFP of PdCoO$_2$. This distinction between isotropic Cu and anisotropic PdCoO$_2$ is further supported by their thickness-dependent resistivity behavior: in Cu, resistivity remains relatively unchanged with surface orientation, whereas in PdCoO$_2$, it varies significantly depending on crystallographic orientation (see SI-Fig.S6). It should be noted that when the MFP is decomposed into orthogonal components, the value averaged over directions is inherently smaller than the MFP, since components perpendicular to that direction contribute negligibly or vanish entirely.
When the thickness is reduced to 100 nm and 10 nm, PdCoO$_2$ shows a negligible variation in both $\lambda\parallel$ and $\lambda_\bot$, indicating that the boundary scattering has minimal effect on its transport properties even in thin-film forms. In contrast, Cu exhibits a sharp thickness dependence. At 100 nm, the MFP remains nearly identical to the bulk, but at 10 nm, the in-plane MFP is reduced by half, and the out-of-plane MFP decreases significantly below 5 nm. These trends suggest a stronger sensitivity to the boundary scattering in Cu, particularly in confined geometries~\cite{Ok2024}.
To visualize how these changes affect the carrier lifetime, we mapped the thickness-dependent relaxation times onto the Fermi surface at each $k$-point (Fig.~\ref{Fig3}(c); see SI-Fig.S4, SI-Fig.S5). For both materials, the relaxation time distributions in the 100 nm films closely resemble those of the bulk. However, in 10 nm films, new low-relaxation-time regions emerge, especially in Cu, where surface boundary scattering becomes dominant. These regions are visualized as blue areas in the Fermi surface projection.
Interestingly, while Cu shows a significant reduction in relaxation time across all directions at 10 nm, PdCoO$_2$ maintains the same relaxation time at 10 nm, with the high relaxation time along the in-plane direction. This persistency highlights PdCoO$_2$’s intrinsic resistance to the surface boundary scattering, owing to its strongly anisotropic and layered structure. The comparison demonstrates that, unlike Cu, PdCoO$_2$ maintains its favorable transport characteristics even at nanometer-scale thicknesses. These results underscore the potential of PdCoO$_2$ as a robust and scalable interconnect material for the next-generation low-power consumption nanoelectronics.

\begin{figure}[t]
\includegraphics[width=1\linewidth]{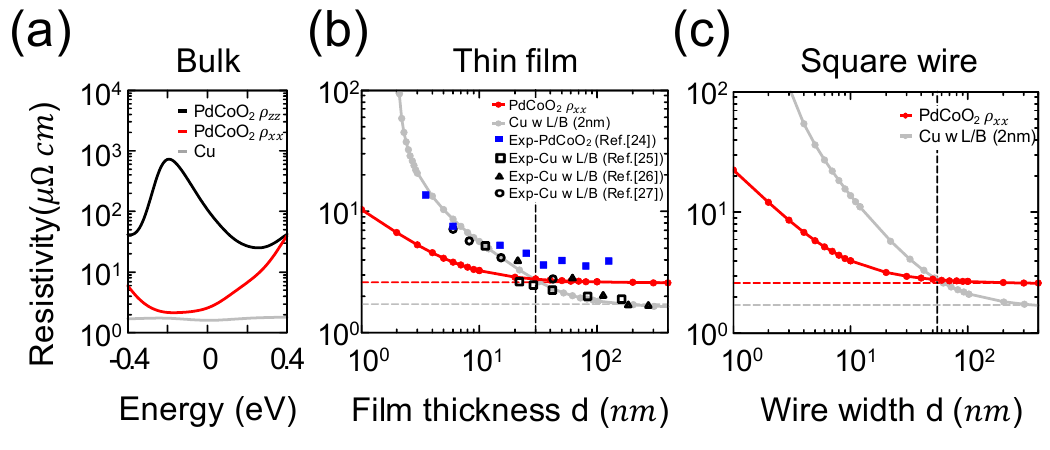}
\caption{Thickness-dependent resistivity of PdCoO$_2$ and Cu at 300~K. (a) Calculated resistivity of bulk PdCoO$_2$ and Cu at 300~K. The red and black lines represent in-plane ($x$) and out-of-plane ($z$)[111] resistivities of PdCoO$_2$, respectively, near the Fermi level. The gray line indicates the isotropic resistivity of bulk Cu. (b) Thickness-dependent resistivity of PdCoO$_2$ and Cu thin films at 300~K. Blue squares show experimental resistivity data for PdCoO$_2$ from a previous experimental study~\cite{Rimal2021}, while black symbols represent Cu data from previous experimental studies~\cite{Gall2020, Chawla2012, Khojier2013}. (c) Width-dependent resistivity of square-shaped PdCoO$_2$ and Cu wires at 300~K. For Cu, a 2-nm-thick liner and diffusion barrier layer was included to reflect realistic interconnect geometries.
\label{Fig4}}
\end{figure}
Finally, we estimate the thickness-dependent conductivity and resistivity of PdCoO$_2$ and Cu by incorporating boundary conditions and geometric effects through a momentum-resolved relaxation time model. Figure~\ref{Fig4}(a) shows the resistivity of the bulk Cu and PdCoO$_2$ at room temperature (300~K), obtained from first-principles calculations based on electron–phonon self-energy. For bulk Cu, the calculated resistivity of 1.59~$\mu\Omega\cdot$cm is in close agreement with the experimental value of 1.73 $\mu\Omega\cdot$cm~\cite{CRC}. For bulk PdCoO$_2$, the calculated in-plane resistivity is 2.7 $\mu\Omega\cdot$cm, which compares well with the experimentally measured value of 2.6 $\mu\Omega\cdot$cm~\cite{Mackenzie2017}. In particular, the bulk PdCoO$_2$ exhibits an extreme anisotropy, with its out-of-plane resistivity more than two orders of magnitude higher than the in-plane value. This agreement indicates that the calculated relaxation times, based solely on electron–phonon scattering, are sufficient to describe the observed resistivity without the need to invoke additional mechanisms such as grain boundary, impurity, or electron–electron scattering. 
Figure~\ref{Fig4}(b) presents the thickness-dependent resistivity of PdCoO$_2$ and Cu thin films predicted by our momentum-resolved model. To simulate realistic interconnect architectures, we incorporate a 2 nm barrier/liner for Cu~\cite{Cuong2017,Li2015,Zhao2014} due to diffusion of Cu~\cite{Lo2017}. The model shows an excellent agreement with the experimental results for both PdCoO$_2$~\cite{Rimal2021} and Cu~\cite{Gall2020, Chawla2012, Khojier2013, Ok2024}. As the thickness of the film decreases, PdCoO$_2$ exhibits a much more gradual increase in resistivity compared to Cu, maintaining values close to the bulk up to 30 nm. At this thickness, the resistivity of PdCoO$_2$ ($\sim 2.7 \mu\Omega\cdot$cm) intersects that of Cu. A detailed analysis of the slope reveals that Cu undergoes a steep resistivity increase around 40 nm, consistent with its isotropic bulk MFP. In contrast, PdCoO$_2$ shows a slope change near 7 nm, governed by its shorter out-of-plane MFP. Our model explains these features well and predicts that resistivity increases sharply when the film thickness becomes comparable to the dominant bulk MFP. For Cu, this occurs at 40 nm; for PdCoO$_2$, due to its anisotropic structure, the out-of-plane MFP leads to a predominant slope change of 7 nm. It should be noted that the theoretical resistivity values predicted for PdCoO$_2$ are slightly lower than the experimental ones indicated by the blue squares in Fig.~\ref{Fig4}(b). This discrepancy arises because our model assumes an ideal defect-free crystal and excludes contributions from grain boundaries (GBs) and other imperfections typically present in experimental samples. The observed gap between theory and experiment, therefore, reflects the additional scattering introduced by GBs, and the magnitude of this deviation may serve as an indirect estimate of GB-related resistivity in the experimental films (see SI-S7 for details).
To further simulate realistic device geometries, we evaluate 1D square wires of Cu and PdCoO$_2$, shown in Fig.~\ref{Fig4}(c). For Cu, the wire geometry introduces additional boundary scattering in both in-plane and out-of-plane directions, resulting in a sharper increase in the resistivity than that in thin films, with a clear slope change around 40 nm. In contrast, PdCoO$_2$ exhibits two distinct transitions: one near 35 nm, where in-plane boundary scattering becomes significant, and another near 7 nm, associated with out-of-plane scattering. This dual slope behavior indicates that anisotropic transport in PdCoO$_2$ leads to a more efficient retention of conductivity in wire geometries. In general, the results demonstrate that PdCoO$_2$ offers superior resistivity scaling in nanoscale interconnects compared to Cu. Its ability to maintain low in-plane resistivity down to tens of nanometers, even in confined geometries, underscores its potential as a high-performance interconnect material for future semiconductor technologies.

\section{\label{sec:level1}Conclusion}
In summary, we have systematically investigated the resistivity behavior of PdCoO$_2$ and Cu using first-principles electron–phonon self-energy calculations combined with a thickness-dependent transport model based on momentum resolution. Our results reveal that PdCoO$_2$ exhibits an extreme anisotropy, with an in-plane resistivity nearly two orders of magnitude lower than its out-of-plane counterpart. This pronounced anisotropy, together with the excellent agreement between theory and experiment, suggests that electron–phonon interactions are the dominant scattering mechanism and that contributions from grain boundaries, defects, and electron–electron interactions can be reasonably neglected in high-quality samples.
We also used a modified relaxation-time formalism derived from the conductivity tensor to evaluate the thickness-dependent resistivity of PdCoO$_2$ and Cu. Compared to Cu, PdCoO$_2$ shows a much slower increase in resistivity with decreasing thickness, retaining values close to the bulk even at sub-30 nm scales. This behavior is consistent with experimental data and reflects the layered structure of PdCoO$_2$ and suppressed boundary scattering in the in-plane direction~\cite{Moll2016}.
The momentum-dependent thickness model also provides a clear framework for interpreting resistivity scaling. As the film thickness approaches the characteristic MFP of each material, the resistivity undergoes a distinct slope change. In Cu, this transition occurs around 40 nm, whereas in PdCoO$_2$, anisotropic MFPs produce multiple transitions: around 35 nm for in-plane confinement and around 7 nm for out-of-plane confinement.
These findings highlight the potential of PdCoO$_2$ as a promising alternative to conventional metals for nanoscale interconnect applications. Its intrinsic anisotropy, low in-plane resistivity, and robust performance under geometric confinement make it an ideal platform for the next-generation electronic devices. Continued research into Pd-based delafossites and related layered oxides may yield innovations in emerging energy-efficient nanoelectronic technologies.


\section*{Acknowledgement} 
This work was supported by the Korean government (MSIT) through the National Research Foundation of Korea (NRF-2022R1A2C1005505 and NRF-2022M3F3A2A01073562) and the Institute for Information \& Communications Technology Planning \& Evaluation (IITP) (2021-0-01580). This work was also supported by the US Department of Energy (DOE), Office of Science, Quantum Science Center (S.-H.K.) for theoretical study, and by the US Department of Energy, Office of Science, Office of Basic Energy Sciences, Materials Sciences and Engineering Division (M.Y., H.N.L.) for theoretical and experimental insights. Computational work was done using the resources of the KISTI Supercomputing Center (KSC-2022-CRE-0379 and KSC-2023-CRE-0053) and the resources of the Oak Ridge Leadership Computing Facility and the National Energy Research Scientific Computing Center, US Department of Energy Office of Science User Facilities.

\bibliography{reference}

\end{document}